# THE AGE OF THE KIC 7177553 SYSTEM

James MacDonald and D. J. Mullan
Department of Physics and Astronomy, DE 19716, USA

ABSTRACT
KIC 7177553 is a quadruple system containing two binaries of orbital periods 16.5 and 18 d. All components have comparable masses and are slowly rotating, non-evolved stars of spectral type near G2V. The longer period binary is eclipsing with component masses and radii, $M_1 = 1.043 \pm 0.014$ $M_\odot$, $R_1 = 0.940 \pm 0.005$ $R_\odot$, and $M_2 = 0.986 \pm 0.015$ $M_\odot$, $R_2 = 0.941 \pm 0.005$ $R_\odot$. The essentially equal radii measurements are inconsistent with the two stars being on the main sequence at the same age using standard stellar evolution models. Instead a consistent scenario is found if the stars are in their pre-main sequence phase of evolution and have age 33 - 36 Myr. Such a young age has important implications for the detectability of the massive planet indicated by eclipse time variations.

## 1. INTRODUCTION

KIC 7177553 was listed in the catalog Detection of Potential Transit Signals in the First Three Quarters of Kepler Mission Data (Tenenbaum et al. 2012) as a binary system of orbital period 18 d. Armstrong et al. (2014) estimated the temperatures of the primary and secondary components of KIC 7177553 to be $5911 \pm 360$ K and $5714 \pm 552$ K, respectively, and derived a radius ratio of $R_2/R_1 = 0.89 \pm 0.28$. Low-amplitude, periodic eclipse timing variations were interpreted as due to the presence of a super-Jupiter planet in an inclined, eccentric 1.45 yr orbit around the binary (Borkovits et al. 2016).

From analysis of the light curve combined with spectroscopic radial velocity measurements, Lehmann et al. (2016, hereafter L16) determined the primary and secondary masses to be $M_1 = 1.043 \pm 0.014$ $M_\odot$ and $M_2 = 0.986 \pm 0.015$ $M_\odot$, respectively. Of particular note is that this ~6% difference in mass is significantly larger than the 0.1% difference in the measured radii, $R_1 = 0.940 \pm 0.005$ $R_\odot$ and $R_2 = 0.941 \pm 0.005$ $R_\odot$. The precision (of order 1%) with which masses and radii have been determined for the components of KIC 7177553 is noteworthy. The high precision has allowed the uncertainties in effective temperatures to be improved by factors of 3-4 compared to those reported by Armstrong et al. (2014). Moreover, L16 have shown that KIC 7177553 belongs to a class of quadruple spectroscopic binaries with at least one eclipsing binary in the system. Another member related to this class has been identified by Rappaport et al. (2016), who report on a quintuple system containing two eclipsing binaries: but in this case, the masses of the eclipsers are determined with considerably poorer precision (as much as 6-7%). The high precision of the mass and radii determinations in KIC 7177553 enable us to make a significant distinction between stars which have reached on the main sequence and those which are still on pre-main sequence tracks. Such a distinction would not be possible in the present case if the precision of masses and radii were as poor as 6-7%.

AGE DETERMINED FROM THE RADIUS MEASUREMENTS

Using an evolution code which has been described previously (MacDonald & Mullan 2013), models have been computed for stars having the masses of the primary and secondary components of the KIC7177553 eclipsing binary. We have considered models for [Fe/H] = -0.1, 0.0 and 0.1, which spans the range in composition determined for the system's 4 spectral type G stars by L16. We calculated two sets of models that used either the OPAL equation of state (eos) (Rogers & Nayfonov 2002) or the SCVH+Z eos (Saumon, Chabrier, & Van Horn 1995; MacDonald & Mullan 2013). Our results are insensitive to the particular choice of eos and we show results only for the SCVH+Z eos. In figure 1, we show radius plotted against age for [Fe/H] = 0.0 models of masses equal to the mean masses of components of the eclipsing binary, i.e. 1.043 and 0.986 $M_\odot$, respectively. The nearly vertical lines on the left-hand side are pre-main sequence tracks, with radii decreasing on short time-scales. Once a star reaches the main sequence, the radius begins to increase, but at a much slower rate than the decreases in the pre-main sequence phase. Also on the main sequence, the primary component at all times maintains a radius which exceeds the radius of the secondary: the excess in radius (about 6%) is related to the excess of primary mass over secondary mass (about 6%). Also shown by horizontal lines are the observational $1\sigma$ limits on the radii. It is clear that there is no co-eval solution for the age if the stars are on the main sequence. If the stars were in fact to be on the main sequence, these models would predict that the primary has age < 80 Myr and the age of the secondary lies between 2.4 and 2.9 Gyr. Such stars would certainly not be co-eval.

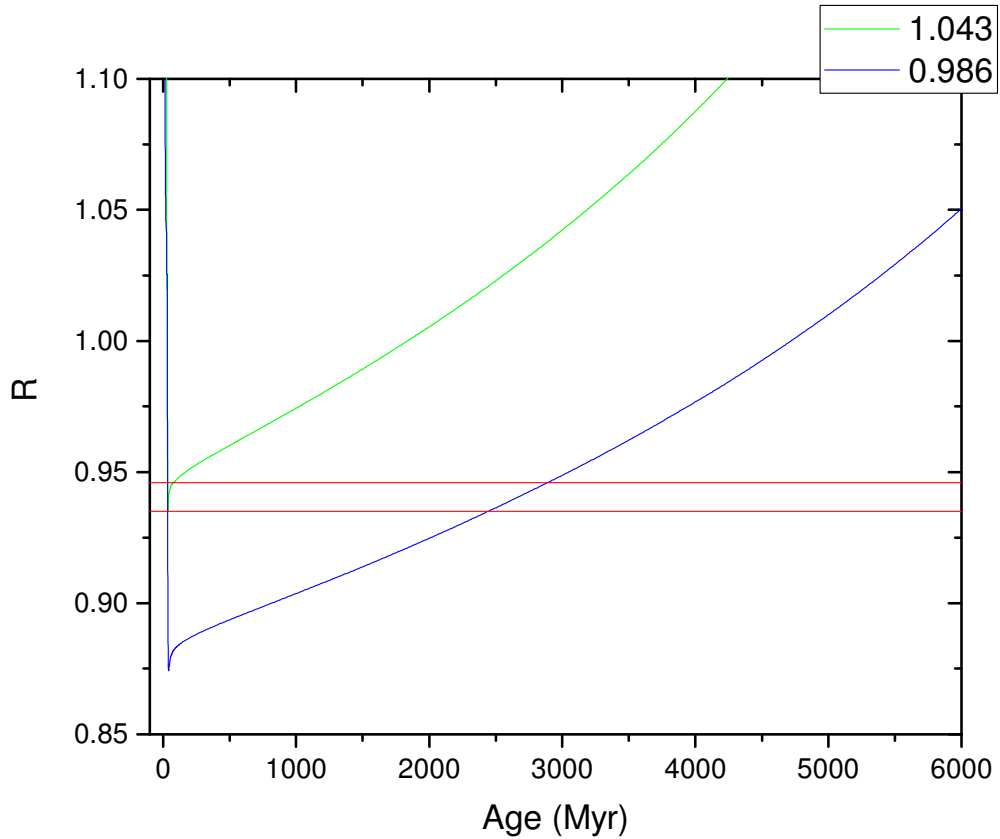

Figure 1. Radius in solar units plotted against age for models with the masses of the components of the KIC7177553 eclipsing binary. The horizontal lines show the $1\sigma$ limits on the measured radii.

Figure 2 shows a portion of the pre-main sequence phase. In this case, it is now possible to see that there are co-eval solutions for the primary and secondary components within a limited range of ages, specifically between ages 33.1 and 33.7 Myr. From spectroscopy, L16 determined the effective temperature, $T_{eff}$, of the primary to be 5800 ±130 K. For the secondary, they found $T_{eff}$ = 5740 ±140 K. These measurements are in excellent agreement with the predictions of our models, $T_{eff}$ = 5840 K and $T_{eff}$ = 5730 K, respectively.

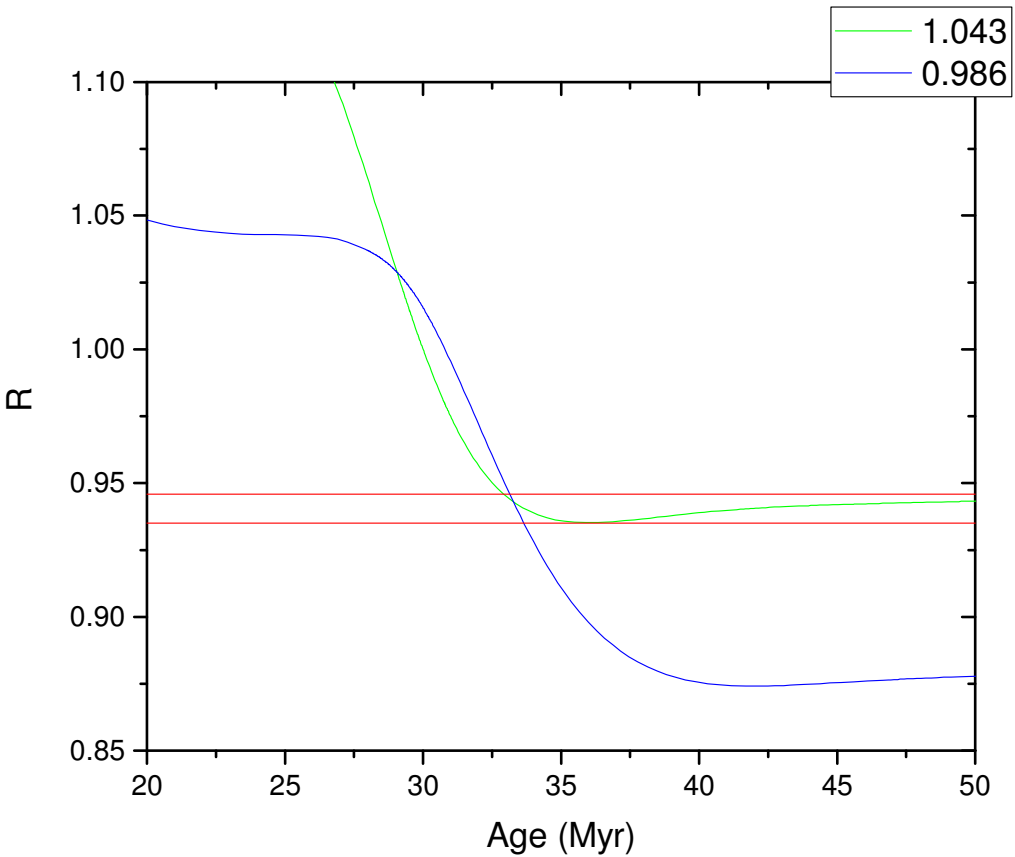

Figure 2. As figure 1, except that the age range is expanded to highlight the evolution of the radii of the primary and secondary components during the pre-main sequence phase: note that both radii fall inside the observed range during a narrowly limited time interval.

We did not find co-eval solutions for models with [Fe/H] = -0.1, because the minimum radius of the model primary was greater than the $1\sigma$ upper limit on the observationally determined radius. The co-eval age from models with [Fe/H] = 0.1 was found to be 35.1 ± 0.3 Myr. The ages determined from the OPAL eos models were found to be slightly higher by ~0.5 Myr.

DISCUSSION

Our modelling suggests that the stars in the KIC 7177553 system are young enough that they are still in the pre-main sequence phase of evolution.

If these were single stars, the expectation would be that at least some of them would be rapidly rotating (e.g. Barnes, Sofia & Pinsonneault, 2001; MacDonald & Mullan 2003). L16 report that the stars in the KIC 7177553 system are slow rotators, which suggests that standard gyrochronology might not be applicable to stars in systems in which a close binary orbits another star or binary system. Other examples of such cases are the two G star + binary brown dwarf systems HD 130948 and Gl 417BC, analyzed by Dupuy, Liu & Ireland (2009) and Dupuy & Liu (2012).

If the stars are not on the pre-main sequence but instead have reached the main sequence, then some physical mechanism not included in standard models is required to inflate the secondary to the same radius as the primary. Mullan & MacDonald (2001) showed that inhibition of convective energy transport due to the presence of a magnetic field does inflate the radius compared to a non-magnetic star but simultaneously makes the star cooler and less luminous. A similar effect is caused by the presence of surface dark spots (Spruit & Weiss 1986; Spruit 1992). In light of the primary and secondary both being slow rotators, we think it unlikely that magnetic fields will be present at the strengths need to inflate the secondary by ~5%. In the spirit of Occam's razor, we consider that a pre-main sequence (non-magnetic) interpretation of KIC 7177553 is more plausible than an interpretation which involves main sequence stars plus magnetic inflation.

If the KIC 7177553 system does contain a super-Jupiter, a young age favors its direct detection by IR observations. Our age estimate of 33 – 36 Myr for the KIC 7177553 system is younger than that of the star HR 8799 which has at least 3 directly imaged planets (Marois et al. 2008). We estimate that at age 33 Myr, a 10 $M_J$ gas giant will have luminosity log $L/L_\odot$ = -4.6 and $T_{eff}$ = 1100 K, which are comparable to the luminosities and temperatures determined by Marois et al. (2008) for HR 8799c and HR 8799d. However, the KIC 7177553 stars are 4.3 magnitudes fainter than HR 8799 in K-band and the orbital period of the planet implies a semi-major axis that is a factor 15 less than that of HR 8799d. Because of its small angular separation from the eclipsing binary, direct imaging of the suspected planet in KIC 7177553 will be challenging, unless it is a highly eccentric orbit.

ACKNOWLEDGMENTS
JM thanks John Gizis for useful discussions.